\documentclass{aa}
\usepackage[varg]{txfonts}

\usepackage{graphicx}
\usepackage{color}

\begin{document}

\title{The Chandra COSMOS Legacy Survey: clustering dependence of Type 2 AGN on host galaxy properties}

\author{V. Allevato\inst{1,2}
\and A. Viitanen \inst{2}
\and A. Finoguenov  \inst{2}
\and F. Civano \inst{3}
\and H. Suh \inst{4}
\and F. Shankar  \inst{5} 
\and A. Bongiorno \inst{6} 
\and A. Ferrara  \inst{1} 
\and R. Gilli \inst{7}
\and T. Miyaji \inst{8}
\and S. Marchesi \inst{7,11}
\and N. Cappelluti \inst{9} 
\and M. Salvato\inst{10}}

\institute{Scuola Normale Superiore, Piazza dei Cavalieri 7, I-56126 Pisa, Italy\\
\email{viola.allevato@sns.it}
\label{inst1}
\and Department of Physics and Helsinki Institute of Physics, Gustaf H\"allstr\"omin katu 2a, 00014 University of Helsinki, Finland
\label{inst2}
\and Harvard-Smithsonian Center for Astrophysics, Cambridge, MA 02138, USA
\label{inst3}
\and Subaru Telescope, National Astronomical Observatory of Japan (NAOJ), 650 North A’ohoku place, Hilo, HI 96720, USA
\label{inst4}
\and Department of Physics and Astronomy, University of Southampton, Highfield, SO17 1BJ, UK
\label{inst5}
\and INAF-Osservatorio Astronomico di Roma, via Frascati 33, 00040, Monteporzio Catone, Italy
\label{inst6}
\and INAF-Osservatorio Astronomico di Bologna, Via Ranzani 1, 40127 Bologna, Italy
\label{inst7}
\and Instituto de Astronomía sede Ensenada, Universidad Nacional Autónoma de México, KM 103, Carret. Tijuana-Ensenada, Ensenada 22860, Mexico
\label{inst8}
\and Physics Department, University of Miami, Knight Physics Building, Coral Gables, FL 33124, USA
\label{inst9}
\and Max Planck-Institute for Extraterrestrial Physics, PO Box 1312, Giessenbachstr. 1., D-85741 Garching, Germany
\label{inst10}
\and Department of Physics and Astronomy, Clemson University, Kinard Lab of Physics, Clemson, SC 29634, USA
\label{inst11}
}

\date{Received Nameofmonth dd, yyyy; accepted Nameofmonth dd, yyyy}

\abstract
    {}
    {We perform clustering measurements of 800 X-ray selected Chandra COSMOS Legacy (CCL) Type 2 AGN with known spectroscopic redshift to probe the halo mass dependence on AGN host galaxy properties, such as galaxy stellar mass M$_{star}$, star formation rate (SFR) and specific black hole accretion rate $\lambda_{BHAR}$, in the redshift range z = [0 - 3]. 
    }
    {We split the sample of AGN with known spectroscopic redshits according to 
    M$_{star}$, SFR and $\lambda_{BHAR}$, while matching the distributions in terms 
    of the other parameters, including redshift.  We measure the projected two-point 
    correlation function $w_p(r_p)$ and model it with the 2-halo term to derive the large-scale 
    bias $b$ and the corresponding typical 
    mass of the hosting halo, for the different subsamples.
    }
    {We found no significant dependence of the large-scale bias and typical halo mass on
    galaxy stellar mass and specific BHAR for CCL Type 2 AGN 
    at mean z$\sim$1, 
    while a negative dependence on SFR is observed, with lower  SFR AGN 
    residing in richer environment. 
    Mock catalogs of AGN matched to have the same X-ray luminosity, stellar mass, $\lambda_{BHAR}$ and SFR of CCL Type 2 AGN, almost reproduce the observed $M_{star}-M_h$, $\lambda_{BHAR}-M_h$ and SFR-M$_h$ relations, when assuming a fraction of satellite AGN $f_{AGN}^{sat} \sim$ 0.15, which corresponds to a ratio between the probabilities of satellite and central AGN of being active $Q \sim$ 2. Mock matched normal galaxies follow a slightly steeper $M_{star}-M_h$ relation - with low mass mock galaxies residing in less massive halos than mock AGN of similar mass, and are less biased than mock AGN with similar specific BHAR and SFR, at least for $Q>$ 1.}
    {}


\titlerunning{AGN clustering dependence on host galaxy properties}
\maketitle

\section{Introduction}

Almost every galaxy in the Universe hosts a super massive black hole (BH) at its center. 
During active phases, when the BH growth is powered by matter accretion, 
the galaxy is observed as an Active Galactic Nucleus
(AGN). BH masses tightly correlate with several properties of their host 
galaxy including stellar mass, velocity
dispersion and galaxy environment. These correlations suggest 
the existence of a fundamental link among BH
growth, host galaxy structure and evolution, and cosmic large-scale 
structure, although the relative importance
of the underlying physical processes are not yet fully understood.

Galaxies, and hence AGN, are not randomly distributed in space. 
On small scales, baryonic matter settles in the potential wells 
of virialized dark matter structures, the so-called halos. On large
scales, the Universe displays coherent structures, with 
groups of galaxies sitting at the
intersections of matter filaments, i.e. “the cosmic web”. 
Clustering is commonly described as the distribution of AGN pairs as
a function of their spatial separation, and it is a 
quantitative measure of the cosmic web topology. It also
provides an indirect measurement of hosting dark matter 
halo masses, statistically classifies the typical AGN
environment, and quantifies how active BHs populate halos. 

Clustering measurements of different types of AGN have 
been carried out by several groups exploiting data from 
multiple surveys in diverse wavebands. Different typical hosting 
halo masses are found in studies at different bands,
ranging from $\sim$10$^{12}$ solar mass for optically selected 
quasars (e.g., Croom et al. 2005, Porciani \& Norberg 2006, Shen et al. 2013; Ross et al. 2009)
to dense environment typical of galaxy groups for 
X-ray selected AGN (Hickox et al. 2009; Cappelluti 
et al. 2010; Allevato et al. 2011; 2012; 2014; Krumpe et al.
2010; Mountrichas et al. 2013; Koutoulidis et al. 2013, Plionis et al. 2018).
However, this difference in the typical halo mass of X-ray compared 
to optically selected AGN may not be present at low redshift (e.g. Krumpe et al. 2012), 
and cannot be explained at present.


\begin{figure}
	\centering
	\includegraphics[width=85mm]{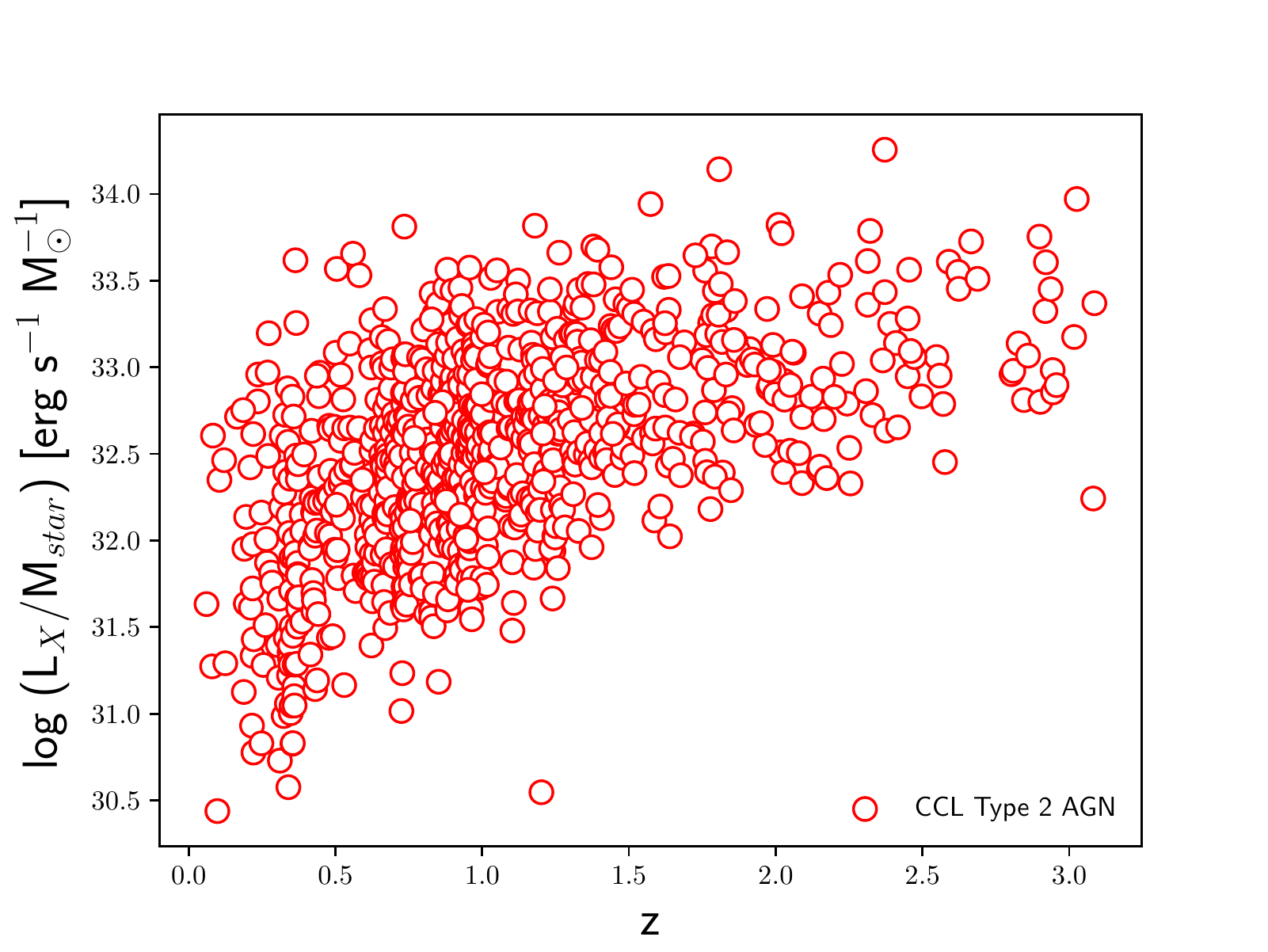}
	\includegraphics[width=85mm]{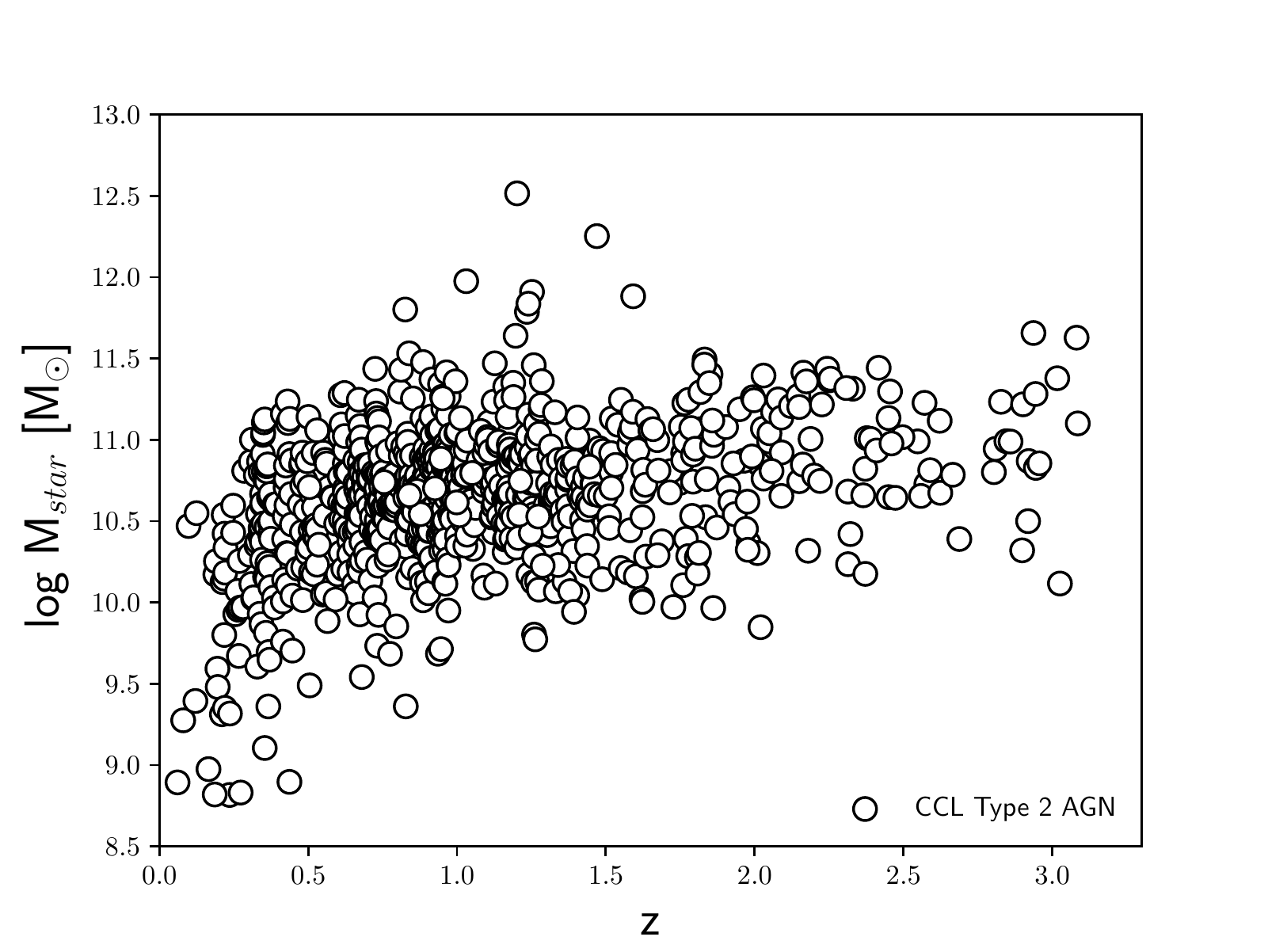}
	\hspace{0mm}
	\centering
	\includegraphics[width=85mm]{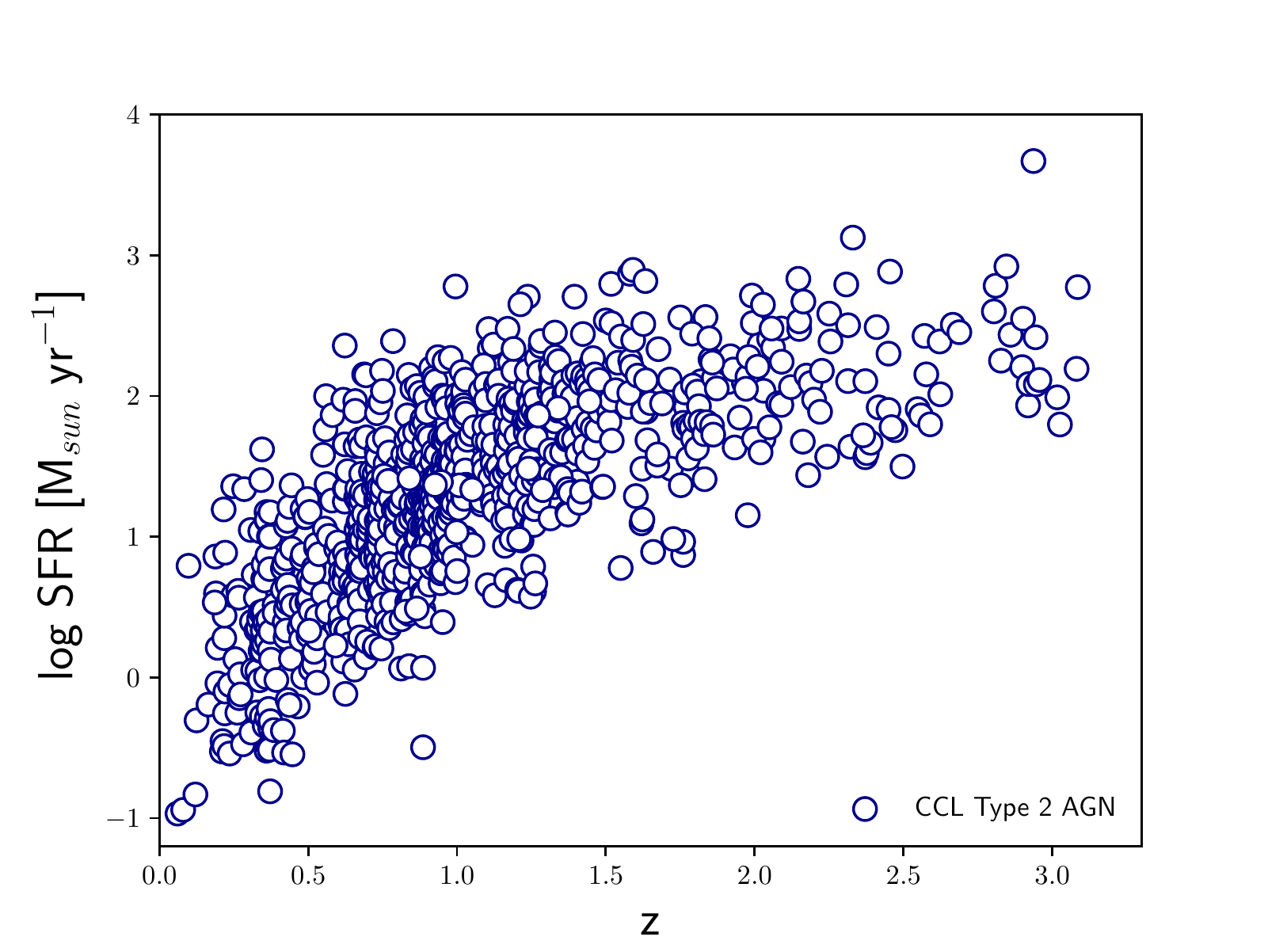}
	\caption{\footnotesize Specific BHAR (Upper Panel), host galaxy stellar mass (Middle Panel) and SFR (Lower Panel) as a function of spectroscopic redshifts for 884 CCL Type 2 AGN with known spectrospic redshifts.}
\end{figure}

Observational biases might be responsible for these different results. 
In fact, recent studies (e.g. Georgakakis et al. 2014; Mendez et al. 2016; Powell et al. 2018, Mountrichas et al. 2019) 
suggest that AGN clustering can be entirely understood in terms of (i) 
galaxy clustering and its dependence on galaxy parameters 
(such as stellar mass and star formation rate), and (ii) AGN selection effects.
In detail, Mendez et al. (2016) compared the clustering of X-ray, radio and 
infrared PRIMUS and DEEP2 AGN with matched galaxy samples 
designed to have the same stellar mass, star-formation rate (SFR), and redshift 
and found no difference in the clustering properties. 
In this scheme, AGN selected using different techniques 
represent separate galaxy populations; the difference in the 
hosting dark matter halos is mainly driven by host galaxy properties. Clustering
studies of large samples of AGN with known host galaxy properties become then crucial
to understand clustering of AGN. 

The clustering dependence on galaxy stellar mass and then the relation between 
stellar/halo mass has been extensively studied during last decade for normal galaxies at 
z$\sim$1 (Zheng et al. 2007; Wake et al. 2011; Meneux et al. 2009; Mostek et al. 2013; Coil et al. 2017)
and at higher redshift (Bielby et al. 2013; Legrand et al. 2018). 
In the sub-halo abundance matching technique, the number density of galaxies 
(from observations) and dark matter halos (from simulations) are matched 
to derive the stellar to halo mass relation at a given redshift (see, e.g., Behroozi et al. 2013, 2018; Moster et al. 2013, 2018, Shankar et al. 2016). 
Moreover, other studies use a halo occupation distribution modeling (Zheng et al. 2007; Leauthaud et al. 2010; Coupon et al. 2015) where a prescription for how galaxies populate dark matter halos can be used to simultaneously predict the number density of galaxies and their spatial distribution. 

The relation between the stellar mass content of a
galaxy and the mass of its dark matter halo is still to investigate 
for active galaxies at all redshift. 
Viitanen et al. (2019) found no clustering dependence on host galaxy 
stellar mass and specific BH accretion rate for a sample of XMM-COSMOS AGN in the range 
z = [0.1 - 2.5]. They also argue that the observed constant halo - galaxy stellar mass relation 
is due to a larger fraction of AGN in satellites (and then in more massive parent halos) in the low $M_{star}$
bin compared to AGN in more massive host galaxies.
Mountrichas et al. (2019) 
suggested that X-ray selected AGN and 
normal galaxies matched to have the same stellar mass, SFR and redshift 
distributions, reside in similar halos and have similar dependence on clustering properties.
They also found a negative clustering dependence on SFR, as also suggested in Coil et al. (2009),
with clustering amplitude increasing with decreasing SFR (see also Mostek et al. 2013 for non active galaxies).

The goal of this work is to extend the clustering measurements performed in Viitanen et 
al. (2019) on XMM-COSMOS AGN to the new Chandra-COSMOS Legacy catalog, building 
one of the largest sample of  X-ray selected AGN detected in a contiguous field. 
This sample of $\sim$ 800 \textit{Chandra} COSMOS Legacy Type 2 AGN 
with available spectroscopic redshift in the range z = [0 -3], 
allows us to study the AGN clustering dependence on host galaxy 
stellar mass, specific BHAR and also SFR while matching 
the distributions in terms of the other parameters, including redshift.

Throughout this paper we use $\Omega_m$ = 0.3, $\Omega_{\Lambda}$=0.7 and $\sigma_8$=0.8 and all distances are measured in comoving coordinates and
are given in units of Mpc h$^{-1}$, where $h = H_0/100 km s^{-1}$.
The symbol log signifies a base-10
logarithm.
In the calculation of the X-ray luminosities we fix $H_0$ = 70 km s$^{-1}$ Mpc$^{-1}$ (i.e. $h$ = 0.7). This is 
to allow comparison with previous studies that also follow similar conventions.

%

\begin{figure*}
	\resizebox{\hsize}{!}{
		\includegraphics[width=0.45\textwidth]{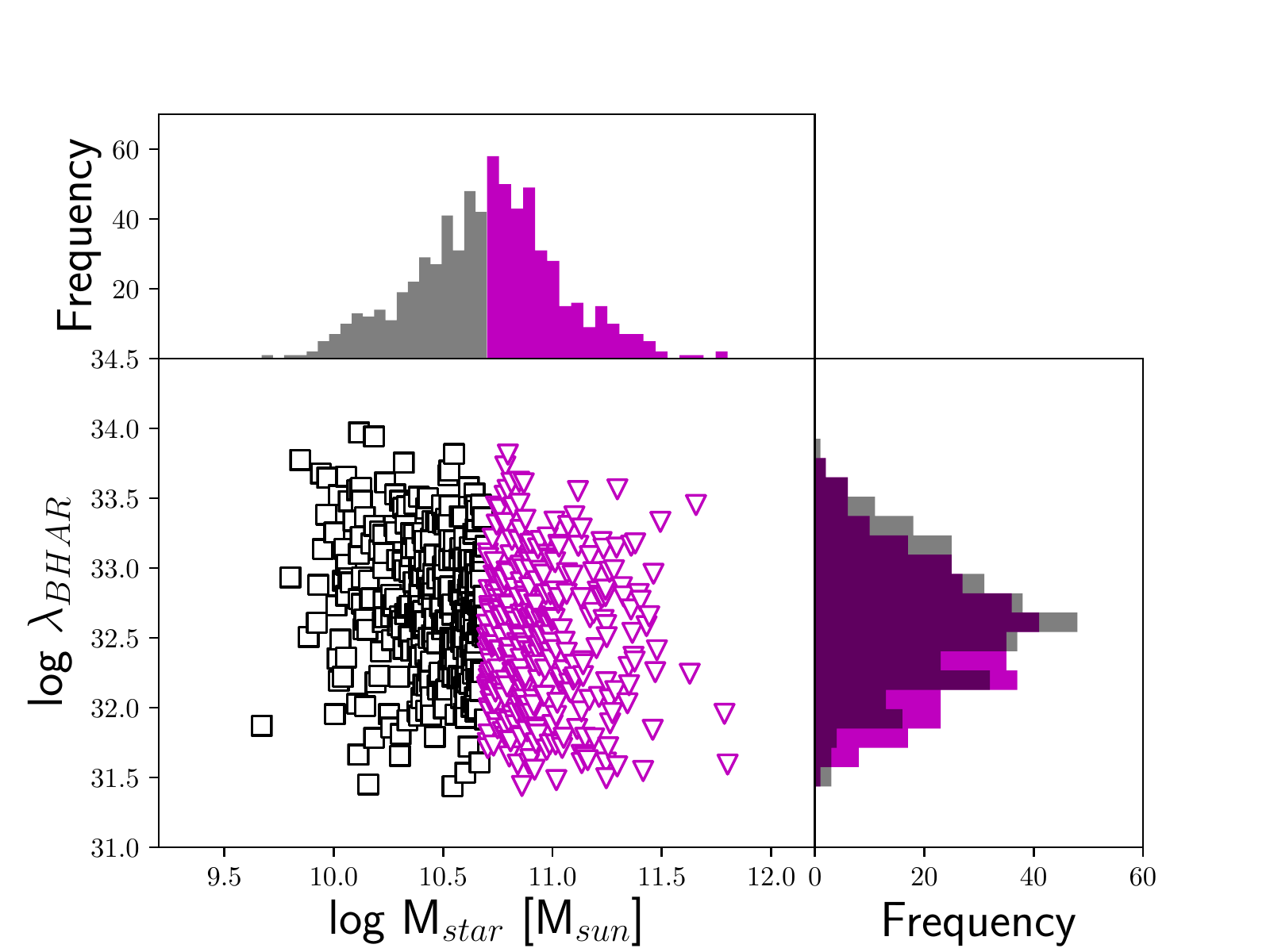}
		\includegraphics[width=0.45\textwidth]{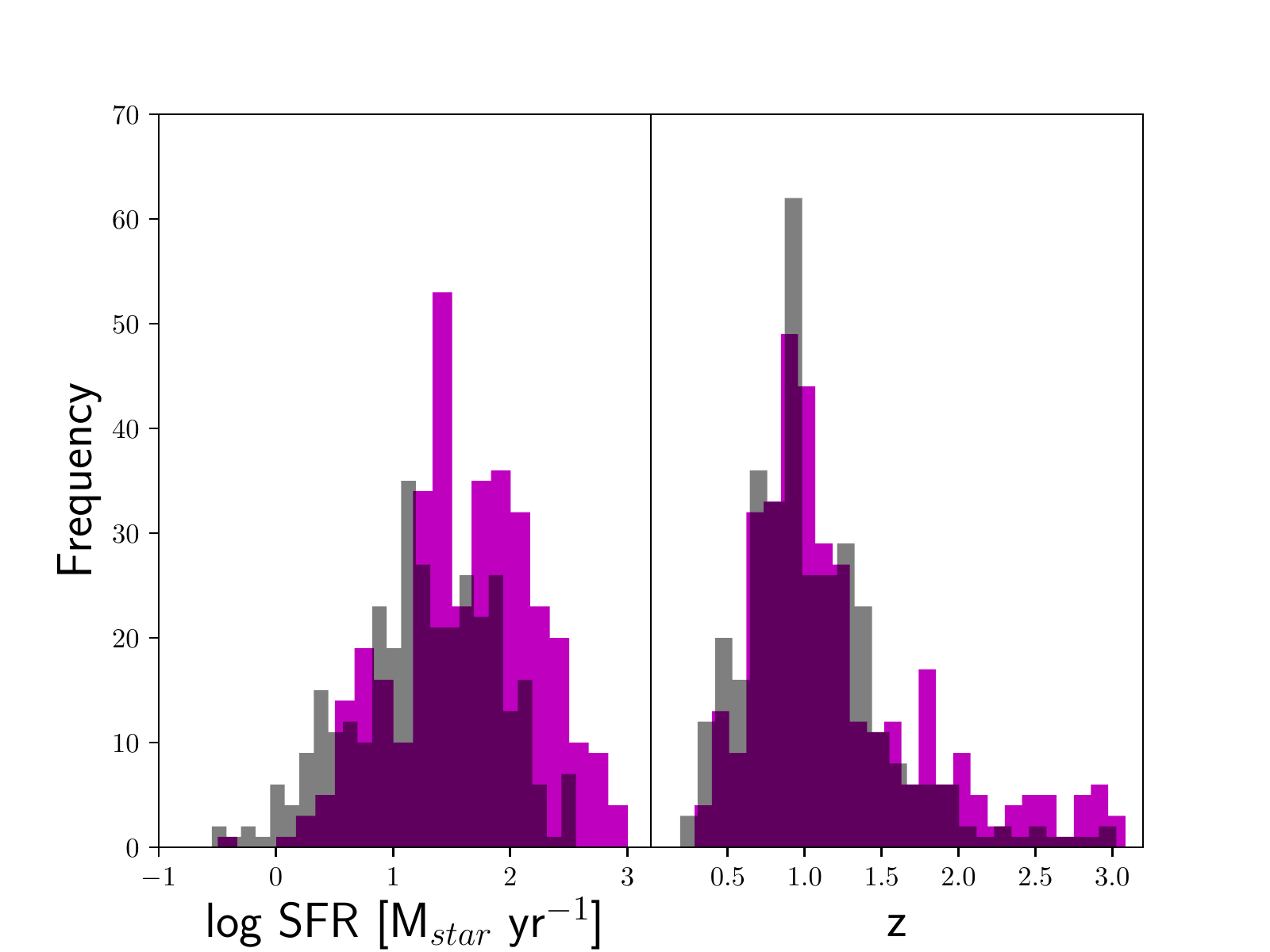}
	}
	\caption{\footnotesize Host galaxy stellar mass as a function of specific BHAR (Left Panel) for \textit{low} (log M$_{star}$/[M$_{\odot}$] $\lesssim$10.75) and \textit{high} ($>$10.75) stellar mass subsamples. The corresponding distribution in terms of M$_{star}$, $\lambda_{BHAR}$, SFR and spectroscopic redshift (Right Panel) are shown for the two AGN subsets.}
	\label{fig:2}
\end{figure*}

\section{Chandra COSMOS Legacy Catalog}

The Chandra COSMOS Legacy Survey CCL (Civano et al. 2016) is a large area, medium-depth X-ray survey covering $\sim$2 deg$^2$ of the COSMOS field obtained by combining the 1.8 Ms Chandra COSMOS survey (C-COSMOS; Elvis et al. 2009) with 2.8 Ms of new Chandra ACIS-I observations. The CCLS is one of the largest samples of X-ray AGN selected from a single contiguous survey region, containing 4016 X-ray point sources, detected down to limiting fluxes of 2.2 $ \times$ 10$^{-16}$ erg cm$^{-2}$ s$^{-1}$, 1.5 $\times$ 10$^{-15}$  erg cm$^{-2}$ s$^{-1}$, and 8.9  $\times$ 10$^{-16}$ erg cm$^{-2}$ s$^{-1}$ in the soft (0.5-2 keV), hard (2-10 keV), and full (0.5-10 keV) bands. 
As described in Civano et al. (2016) and Marchesi et al. (2016), 97\% of CCL sources were identified in the optical and infrared bands and therefore photometric redshifts were computed. Thanks to the intense spectroscopic campaigns in the COSMOS field, $\sim$54\% of the X-ray sources have been spectroscopically identified and classified. The full catalog of CCLS has been presented by Civano et al. (2016) and Marchesi et al. (2016), including X-ray and optical/infrared photometric and spectroscopic properties.

The host galaxy properties of 2324 Type 2 CCL AGN have been studied in the redshift range z = [0 - 3] in 
Suh et al. (2017, 2019). These sources are classified as non-broad-line and/or obscured AGN (here-after, “Type 2” AGN), i.e., they show only narrow emission-line and/or absorption-line features in their spectra, or their photometric spectral energy distributions (SED), is best fitted by an obscured AGN template or a galaxy template.
Making use of the existing multi-wavelength photometric data available in the COSMOS field, they performed a multi-component modeling from far-infrared to near-ultraviolet using a nuclear dust torus model, a stellar population model and a starburst model of the SEDs. Through detailed analyses of SEDs, they derived stellar masses in the range 
9 $<$ log M$_{star}$/M$_{\odot}$ $<$ 12.5 with uncertainties of $\sim$0.19 dex. 
Moreover, SFR are estimated by combining the contributions from UV and
IR luminosity. The total sample spans a wide range of SFRs (-1 $<$ log SFR [M$_{\odot}$ yr$^{-1}$] $<$ 3.5 with uncertainties of $\sim$0.20 dex.

For this study we select 1701/2324 CCL Type 2 AGN detected in the soft band, and focus on 884/1701 sources with known spectroscopic redshift up to z = 3.
%
\begin{figure*}
	\resizebox{\hsize}{!}{
		\includegraphics[width=0.45\textwidth]{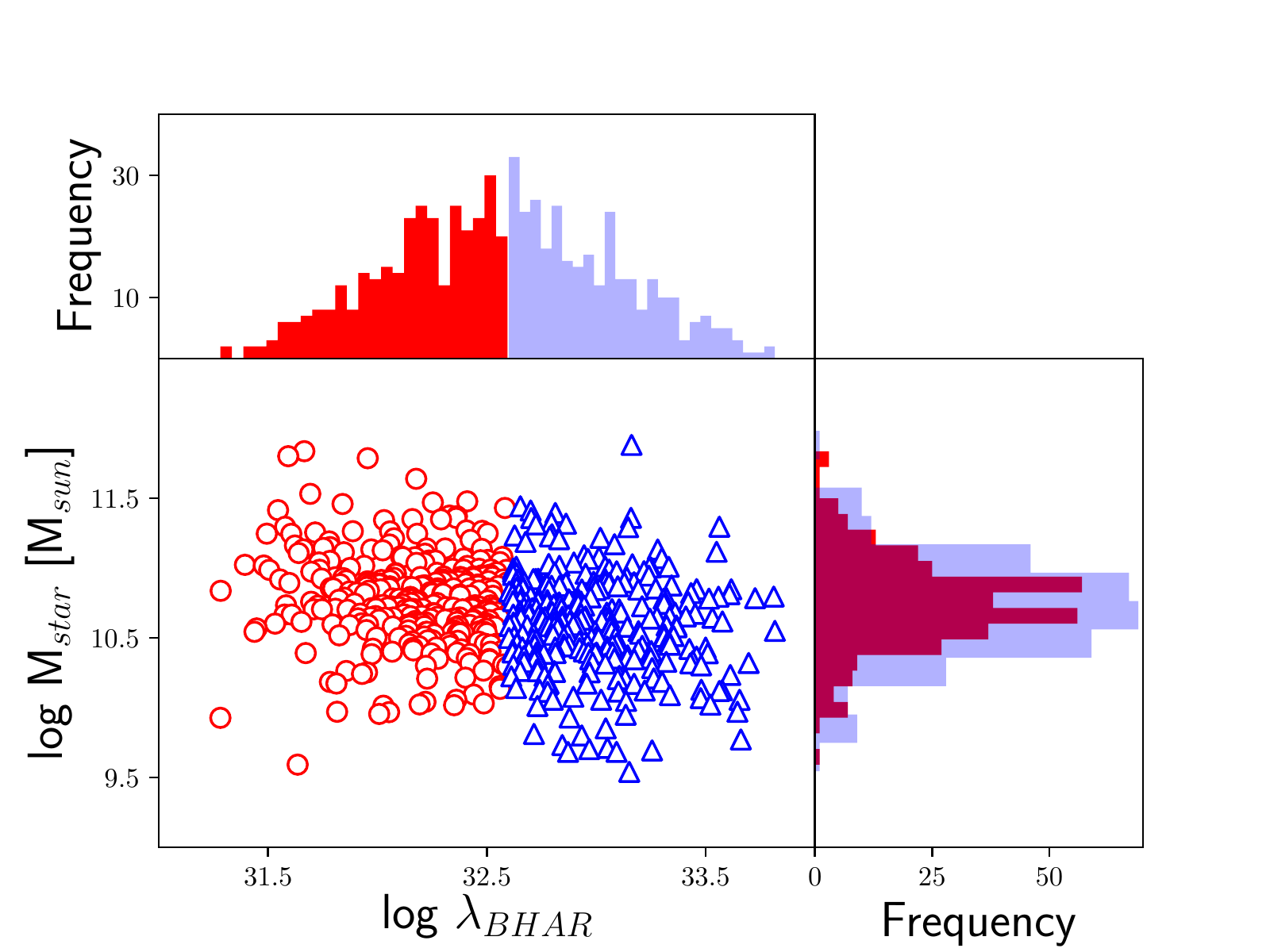}
		\includegraphics[width=0.45\textwidth]{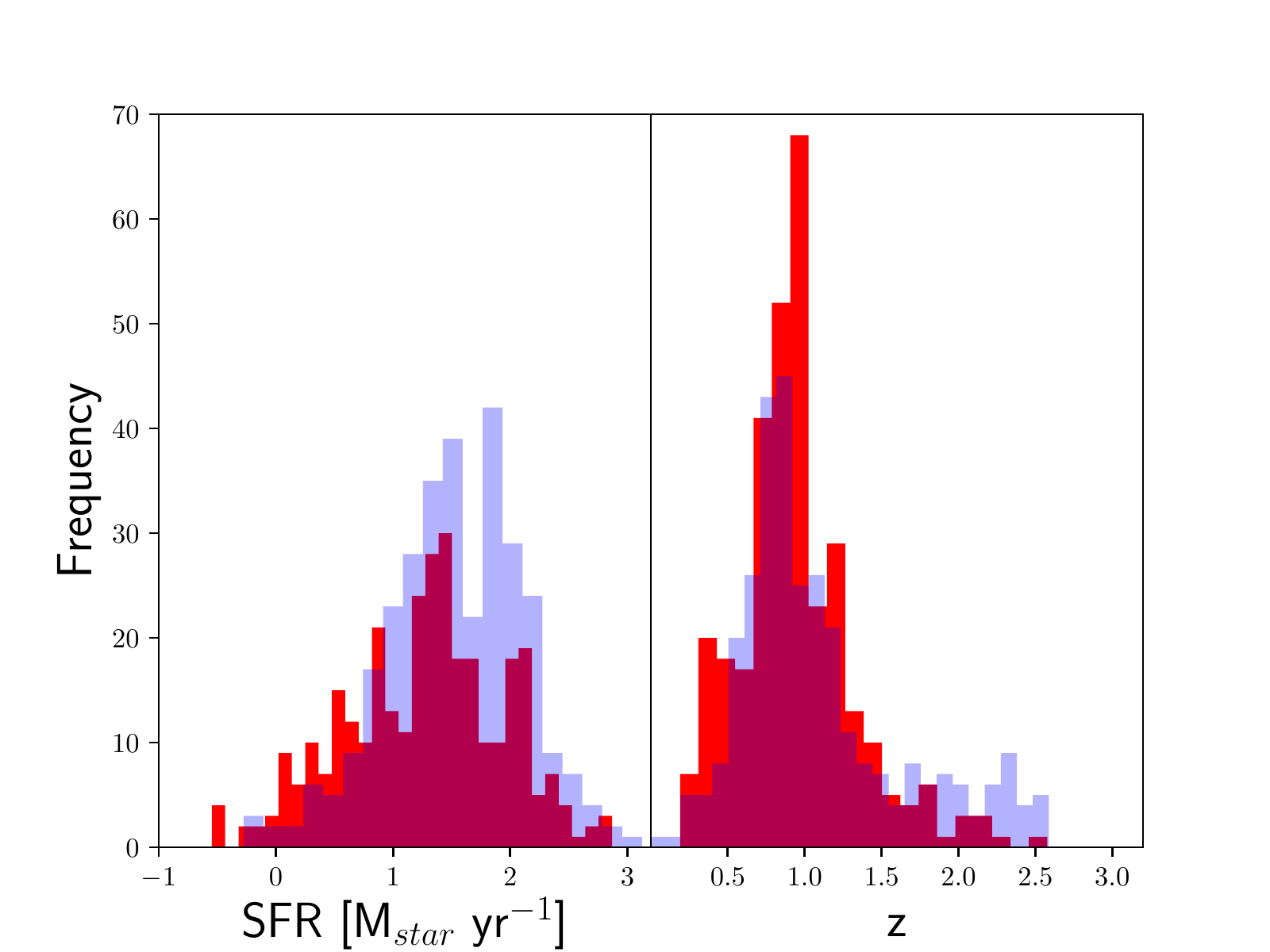}
	}
	\caption{\footnotesize Specific BHAR as a function of host galaxy stellar mass (Left Panel) for \textit{low} (log $\lambda_{BHAR}$ $\lesssim 32.6$) and \textit{high} ($>32.6$) specific BHAR subsamples. The corresponding distribution in terms of M$_{star}$, $\lambda_{BHAR}$, SFR and spectroscopic redshift (Right Panel) are shown for the two AGN subsets.}
	\label{fig:3}
\end{figure*}
%
\begin{figure*}
	\resizebox{\hsize}{!}{
		\includegraphics[width=0.45\textwidth]{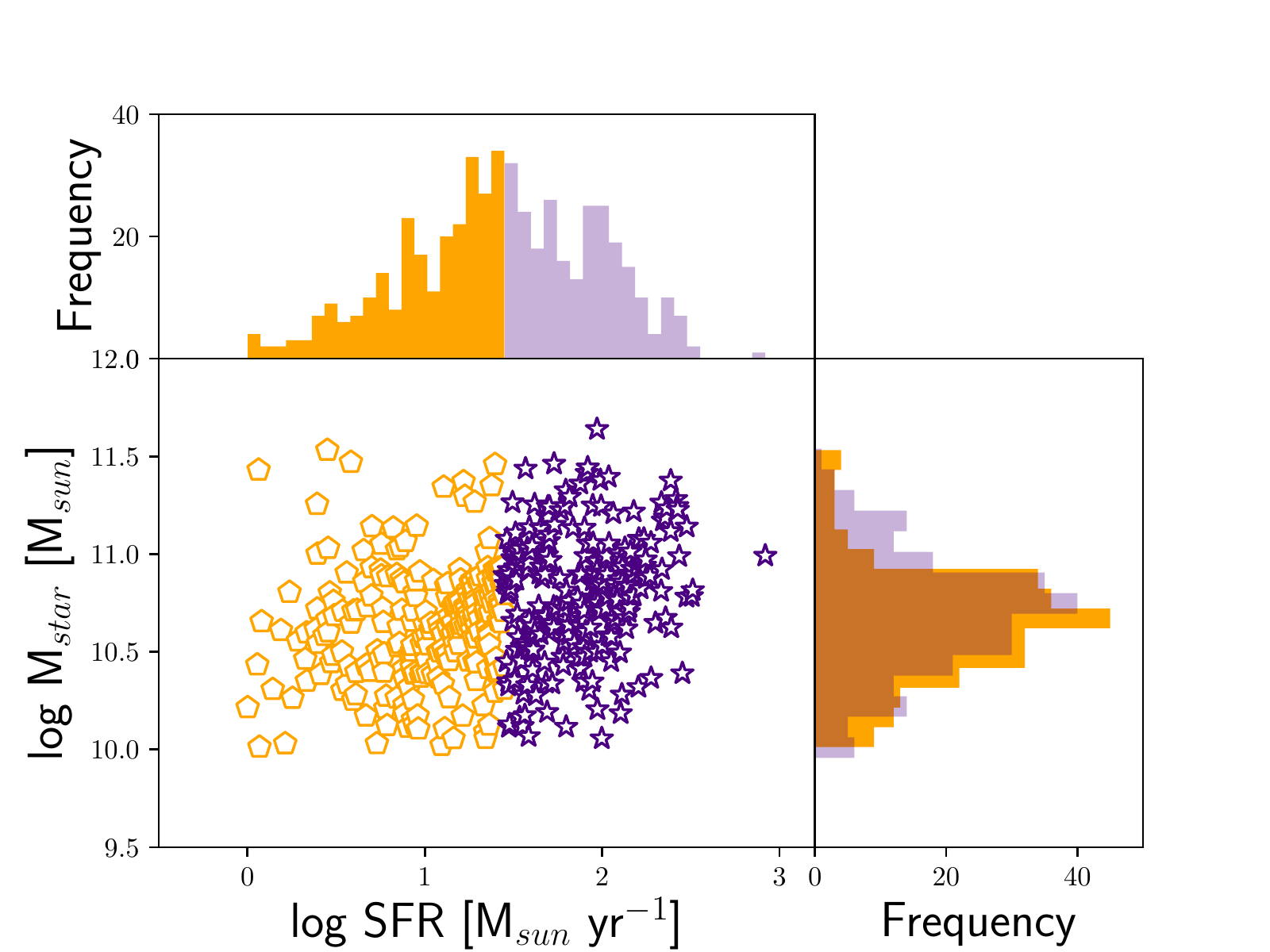}
		\includegraphics[width=0.45\textwidth]{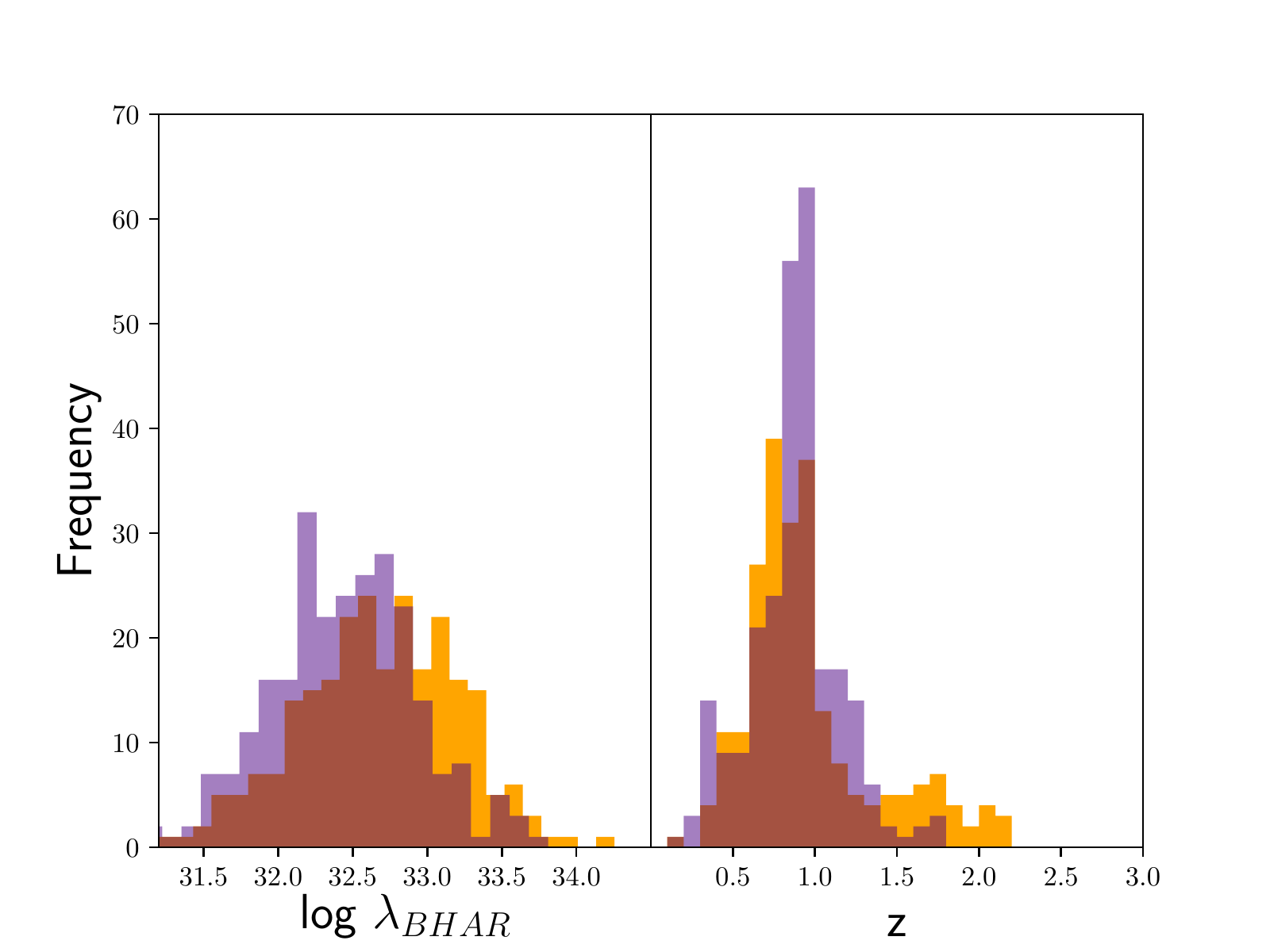}
	}
	\caption{\footnotesize SFR as a function of host galaxy stellar mass (Left Panel) for \textit{low} (log SFR $\lesssim 1.4$) and \textit{high} ($>1.4$) SFR subsamples. The corresponding distribution in terms of M$_{star}$, $\lambda_{sBHAR}$, SFR and spectroscopic redshift (Right Panel) are shown for the two subsets}
	\label{fig:2}
\end{figure*}
In order to study the AGN clustering dependence on host galaxy properties we  divided the sample according to the galaxy stellar mass, SFR and specific 
black hole accretion rates (BHAR) $\lambda_{BHAR} = L_X/M_{star}$. The latter defines the rate of accretion
onto the central BH scaled relative to the stellar mass of the host galaxy. 
To the extent that a proportionality between the BH mass and the host galaxy mass can be assumed,
this ratio gives a rough measure of the Eddington ratio.
Following Bongiorno et al. (2012, 2016) and Aird et al. (2018):
\begin{equation}
\lambda_{Edd} = \frac{k_{bol} \centerdot A} {1.3 \times 10^{38}} \times \frac{L_X}{M_{star}}
\end{equation}
where $L_X$ is the intrinsic X-ray luminosity in erg s$^{-1}$, $k_{bol}$ is a bolometric 
correction factor and 
$M_{star}$ is the total stellar mass of the AGN host galaxy in units of $M_{\odot}$.
The factor $A$ is a constant 
if the BH mass can be related to the host galaxy mass through scaling relations (with A $\thickapprox$ 500-1000; Magorrian et al. 1998; Haring \& Rix 2004). 
Thus, for a mean bolometric correction of $k_{bol}$ = 25 and a constant 
host stellar to black hole mass ratio of A = 500, a ratio of $L_X/M_{star}$ = $10^{34}$  [erg s$^{-1}$ M$_{star}$]
would approximately correspond to the Eddington limit.
The specific BHAR distribution can then be regarded as a tracer of the distribution of Eddington ratios. 
For the porpose of our study, with a fixed k$_{bol}$, $\lambda_{Edd} \propto \lambda_{BHAR}$.
Figure 1 shows the distribution of the specific BHAR, host galaxy stellar mass and SFR for the sample of 884 CCL Type 2 AGN with known spectroscopic redshift in the range z = [0 - 3]. 

We then made subsamples in bins of galaxy stellar mass, specific BHAR and SFR.
Specifically, to avoid selection effects between the different bins of stellar mass (see Powell et al. 2018)
we defined 2 bins of $M_{star}$
and then for each bin, we randomly selected $N$ AGN of the sample,
with $N$ the larger number of sources in the bin, to match the distributions in redshift, 
$\lambda_{BHAR}$ and SFR. Similarly, we defined 2 bins in specific BHAR (SFR) and 
randomly selected subsamples with similar redshift, stellar mass and SFR ($\lambda_{BHAR}$) distributions. 
The final subsamples consist of:
(a) 362 \textit{low} and 374 \textit{high} galaxy stellar mass AGN by using a cut at 
log (M$_{star}$/M$_{\odot}$)  $=$ 10.7 (see Figure 2);
(b) 339 \textit{low} and 326 \textit{high} specific BHAR AGN with a cut at 
log ($\lambda_{BHAR}$/[erg s$^{-1}$ M$_{star}^{-1}$]) $=$ 32.6 (see Figure 3);
(c) 262 \textit{low} and 247 \textit{high} SFR AGN cutting at 
log (SFR/M$_{star}$ yr$^{-1}$)  $=$ 1.4 (see Figure 4).
The characteristics of these subsamples are
summarized in Table 1.


\section{Projected 2pcf}

The most commonly used quantitative measure of large scale structure
is the 2pcf, $\xi$(r), which traces the amplitude of AGN clustering as a function of scale. $\xi$(r) is defined as a measure of the excess probability $dP$, above what is expected for an unclustered random Poisson distribution, of finding an AGN in a volume element $dV$ at a separation $r$ from
another AGN:
\begin{equation}
dP = n[1 + \xi(r)]dV
\end{equation}
where $n$ is the mean number density of the AGN sample 
(Peebles 1980). Measurements of $\xi$(r) are generally performed in comoving space, with
$r$ having units of h$^{-1}$ Mpc.

We measured the projected 2pcf in bins of r$_p$ and $\pi$ (distances perpendicular and parallel to the line of sight, respectively) using CosmoBolognaLib, a large set of Open Source C++ numerical libraries for cosmological calculations (Marulli et al. 2016), which counts the number of pairs of galaxies in a catalog separated by r$_p$ and $\pi$. We then projected through redshift space to eliminate any redshift-space
and we estimate the so-called projected correlation function $w_p(r_p)$ \citep{Dav83}:
\begin{eqnarray}\label{eq:integral}
w_p(r_p) = 2 \int_0^{\pi_{max}} \xi(r_p,\pi) d\pi 
\end{eqnarray}

where $ \xi(r_p,\pi)$ is the two-point correlation function in terms of
$r_p$ and $\pi$, measured using the \citet[LS]{Lan93} estimator:
\begin{equation}\label{eq:LZ}
\xi = \frac{1}{RR'} [DD'-2DR'+RR']
\end{equation}
where DD', DR' and RR' are the normalized data-data, data-random and random-random pairs.

The measurements of the 2pcf requires the construction of a random 
catalog with the same selection criteria and observational effects as the data.
To this end, we constructed a random catalog 
where each simulated source is placed at a random position in the sky, with
its flux randomly extracted from the catalog of real source fluxes.
The simulated source is kept in the random sample 
if its flux is above the sensitivity map value at
that position \citep{Miy07, Cap09}.
The corresponding redshift for each random object is then assigned based on the
smoothed redshift distribution of the AGN sample. 

The value of $\pi_{max}$ is chosen such that the amplitude of the 
projected 2pcf converges and gets noisier for any higher values. 

We calculated the covariance matrix via the jackknife resampling method:
\begin{eqnarray}
C_{i,j} = \frac{M}{M-1} \sum_{k}^{M} \left[ w_k (r_{p,i}) - \langle w(r_{p,i}) \rangle \right]  \times  \nonumber\\
\times  \left[ w_k(r_{p,j}) - \langle w(r_{p,j}) \rangle \right] 
\end{eqnarray}

where we split the sample into M = 9 sections of the sky, and computed the cross-correlation function when excluding each section ($w_k$). We quote the errors on our measurement as the square root of the diagonals, $\sigma_i = \sqrt{C_{i,i}}$.

In the halo model approach, the large scale amplitude
signal is due to the correlation between objects in distinct halos
and the bias parameter defines the 
relation between the large scale clustering amplitude 
of the AGN correlation function and the DM 2-halo term:
\begin{equation}\label{eq:b}
b^{2-h}(r_p)=(w_{AGN}(r_p)/w_{DM}^{2-h}(r_p))^{1/2}
\end{equation}
We first estimated the DM 2-halo term at the median redshift of the sample, using:
\begin{equation}
w_{DM}^{2-h}(r_p)=r_p \int_{r_p}^{\infty} \frac{\xi^{2-h}_{DM}(r)rdr}{\sqrt{r^2-r_p^2}}
\end{equation}
where
\begin{equation}\label{eq:2-halo}
\xi^{2-h}_{DM}(r)=\frac{1}{2\pi^2}\int P^{2-h}(k)k^2 \left[ \frac{sin(kr)}{kr} \right]  dk
\end{equation}
$P^{2-h}(k)$ is the Fourier Transform of the linear power spectrum. In particular, we base our estimation of the linear power spectrum on Eisenstein \& Hu (1999), which is also implemented in CosmoBolognaLib.

\begin{figure}
	\centering
	\includegraphics[width=85mm]{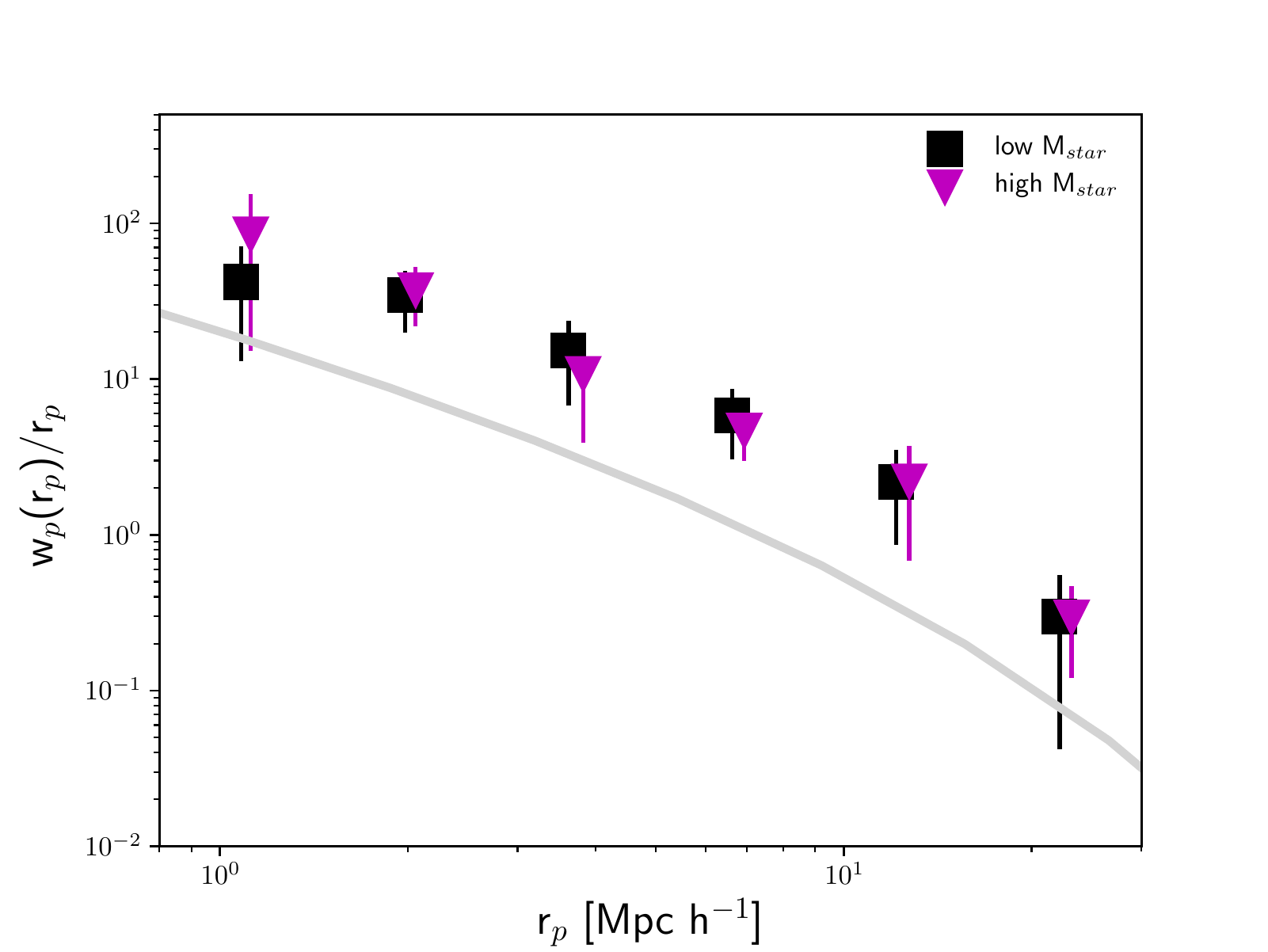}
	\includegraphics[width=85mm]{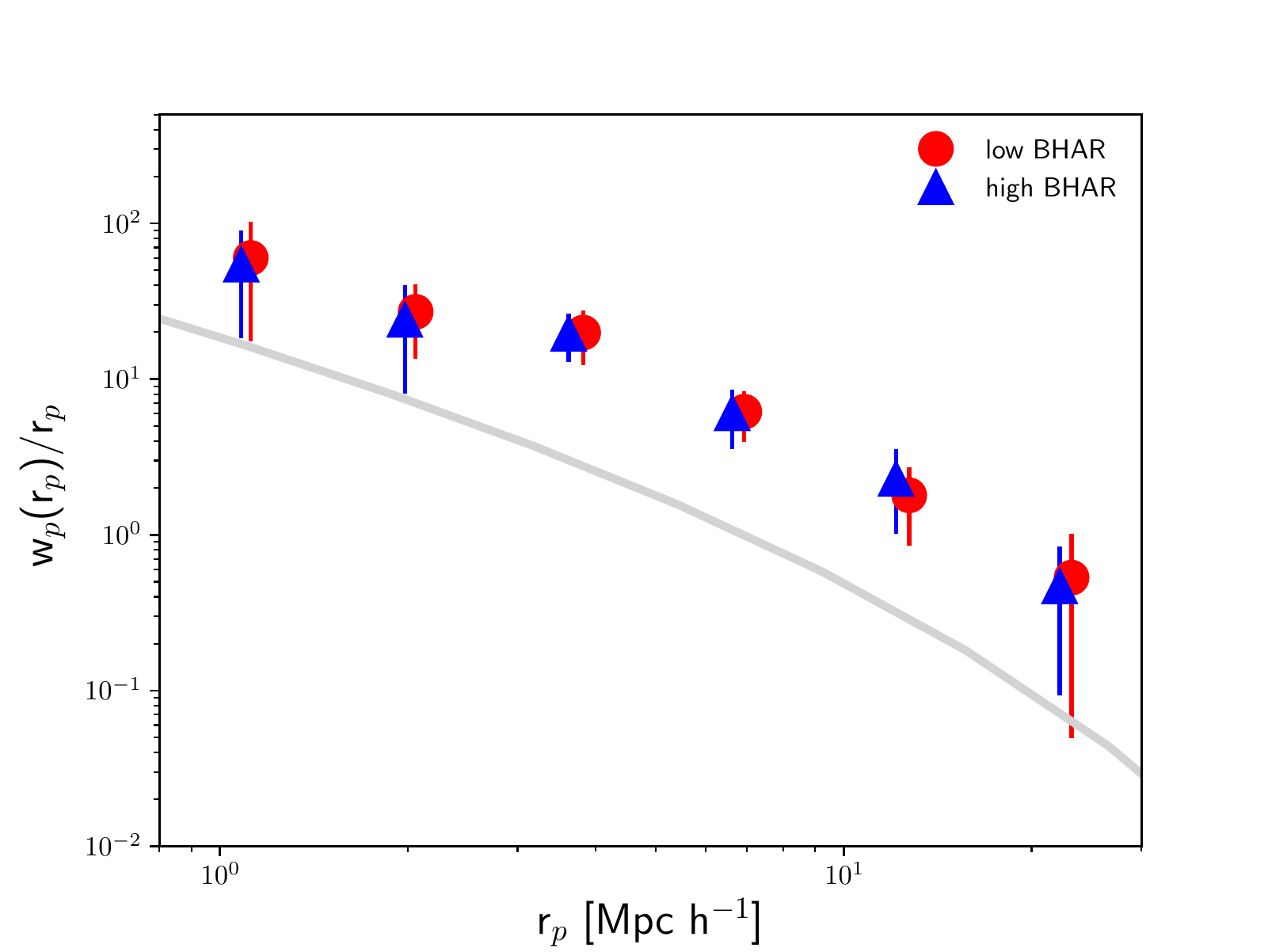}
	\hspace{0mm}
	\centering
	\includegraphics[width=85mm]{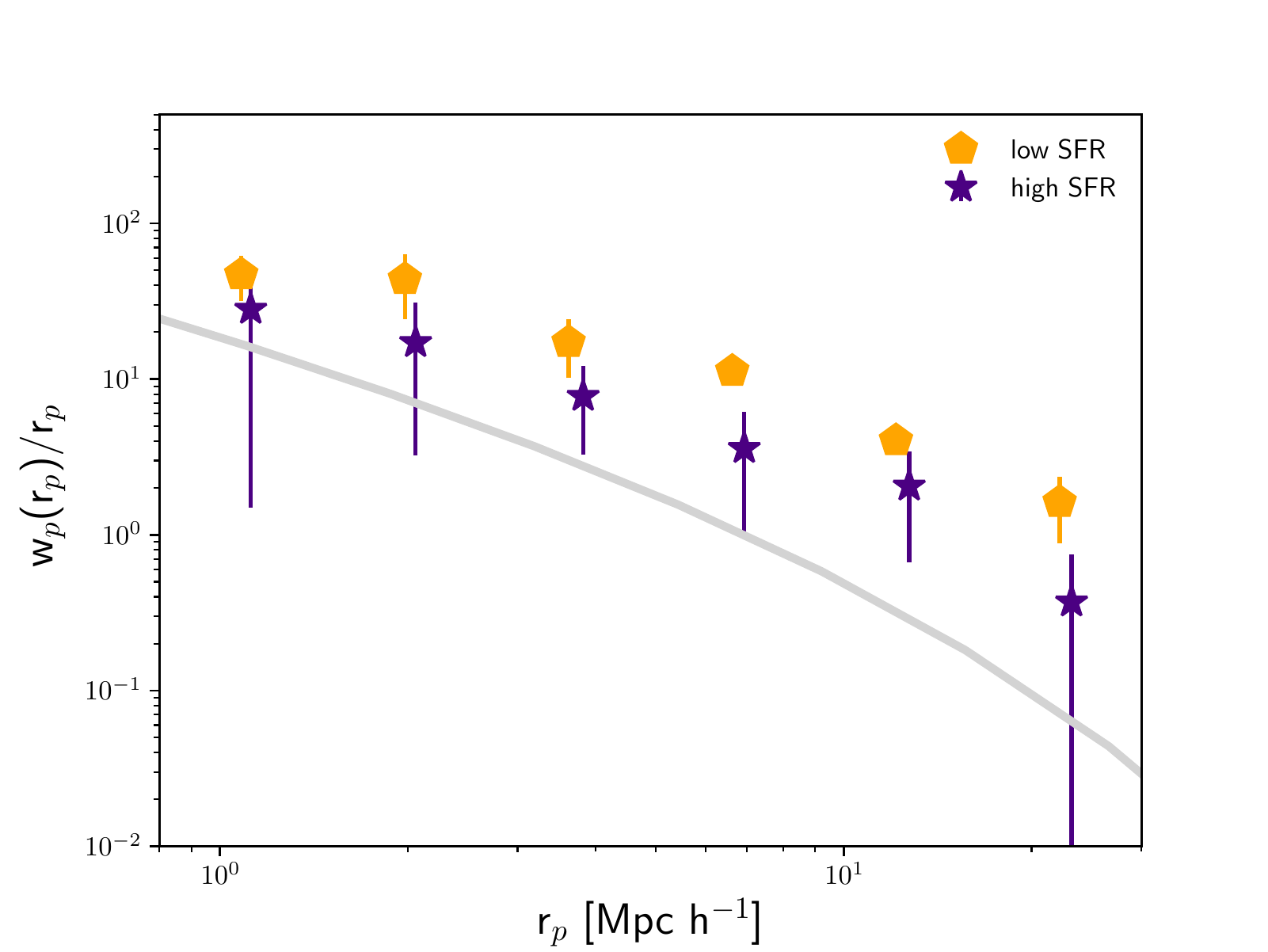}
	\caption{\footnotesize Projected 2pcf of CCL COSMOS Type 2 AGN as a function of: (a) host galaxy stellar mass, logM$_{star}$ $<$ 10.7 (black squares) and $\gtrsim$ 10.7 (magenta triangles); (b) specific BHAR, log$\lambda_{BHAR}$ $<$ 32.6 (red circles) and $\gtrsim$ 32.6 (blue triangles); (c) logSFR $<$ 1.4 (orange pentagons) and $\gtrsim$ 1.4 (purple stars). The grey line shows the DM projected 2pcf at mean z $\sim$ 1.}
\end{figure}


%
%

\begin{table*}[htbp]
	\centering
	\caption{Properties of the AGN Samples.}
	\begin{tabular}{lcccccccr}
		& Sample
		& $z$
		& log L$_{X}/M_{star}$
		& log $M_{star}$
		& log SFR
		& $b$
		& log M$_{typ}$ \\
		& {}
		& median
		& erg s$^{-1}/M_{\odot}$
		&$M_{\odot}$
		&$M_{\odot}$ yr$^{-1}$
		&{}
		&h$^{-1}$ M$_{\odot}$\\		
		\hline
		\\
		Low M$_{star}$ & 362 & 0.96   & 32.7  & 10.4   & 1.4 & 2.14$^{+0.08}_{-0.08}$  & 12.93$^{+0.06}_{-0.06}$ \\		
		High M$_{star}$ & 374 & 1.01    & 32.5   & 10.9   & 1.6 & 2.31$^{+0.09}_{-0.09}$ & 13.03$^{+0.07}_{-0.07}$\\	
		Low $\lambda_{BHAR}$ & 300  & 0.91  & 32.2  & 10.7   & 1.4 & 2.32$^{+0.11}_{-0.11}$ & 13.02$^{+0.08}_{-0.08}$ \\		
		High $\lambda_{BHAR}$ & 306& 1.03  & 32.9  & 10.6  & 1.5 &2.40$^{+0.10}_{-0.10}$ & 13.11$^{+0.06}_{-0.06}$\\		
		Low SFR & 251 & 0.92   & 32.5 & 10.7   & 1.0 & 2.43$^{+0.06}_{-0.06}$  & 13.14$^{+0.06}_{-0.06}$ \\		
		High SFR & 260 & 1.09 & 32.6  & 10.6   & 1.9 & 2.16$^{+0.08}_{-0.08}$ &12.94$^{+0.06}_{-0.06}$\\
		
	\end{tabular}
	\label{tab:results}
\end{table*}

\section{Results}

\subsection{AGN bias versus M$_{star}$}

The goal of this paper is to study the AGN clustering dependence 
on host galaxy stellar mass, specific BHAR and SFR. 
In this section, we focus on the clustering properties of CCL Type 2 COSMOS AGN 
as a function of host stellar mass M$_{star}$.

As shown in Figure 5 (upper panel), the projected 2pcf $w_p$($r_p$) of 362 (374) AGN 
with log (M$_{star}$/M$_{\odot}$)  $<$ 10.7 ($\gtrsim$ 10.7) has been measured 
in the r$_p$ range 1-30 h$^{-1}$ Mpc, following Equation (3). 
The typical value of $\pi_{max}$ used in clustering measurements of
both optically selected luminous quasars and X-ray selected
AGN is $\sim$ 20 - 100 h$^{-1}$ Mpc (e.g., Zehavi et al. 2005; Coil
et al. 2009; Krumpe et al. 2010; Allevato et al. 2011). We found the
optimal $\pi_{max}$ value to be = 60 h$^{-1}$ Mpc, by deriving the 
value at which the amplitude of the signal appears to level off.

The 1$\sigma$ errors on $w_p$($r_p$) are the square root of
the diagonal components of the covariance matrix (Miyaji
et al. 2007; Krumpe et al. 2010, Allevato et al. 2016) estimated using the jackknife method. 
The latter quantifies the level of correlation between
different bins.

Following Equation (6), we derive the best-fit bias by using a
$\chi^2$ minimization technique with one free parameter $\chi^2$ = $\Delta^{T}M^{-1}_{cov}\Delta$. In detail, $\Delta$ is a vector composed of $w_p$($w_p$) - $w_{mod}$($r_p$), $\Delta^{T}$ is its transpose and $M^{-1}_{cov}$ is the inverse of covariance
matrix. The latter full covariance matrix is used in the fit to take into account the correlation between errors.

As shown in Table 1, we derived for the 
\textit{low} and \textit{high} M$_{star}$ subsamples b = 2.14$^{+0.08}_{-0.08}$ and b = 2.31$^{+0.09}_{-0.09}$
at mean z $\sim$ 1, respectively.
Following the bias-mass relation b(M$_h$, z) described in van den Bosch (2002)
and Sheth et al. (2001), the large-scale bias values correspond to typical
masses of the hosting halos of log(M$_h$/M$_{\odot}$ h$^{-1}$) = 12.93$^{+0.06}_{-0.06}$ and 13.03$^{+0.07}_{-0.07}$, for the \textit{low} and \textit{high} stellar mass subsamples, respectively. 
It is worth reminding that these two AGN subsets have similar distributions in terms 
of specific BHAR, SFR and redshift.
Through this work, we refere to
typical halo mass as the DM halo mass which
satisfies b = b(M$_{halo}$) (e.g. Hickox et al. 2009; Allevato et al. 2014, 2016; 
Mountrichas et al. 2019).

\subsection{AGN bias versus $\lambda_{BHAR}$ and SFR}

In this section  we investigate the clustering dependence of CCL
Type 2 COSMOS AGN on specific BHAR and SFR. 
For this purpose, we estimated the projected 2pcf of: 
(a) 300 \textit{low} and 306 \textit{high} 
specific BHAR AGN by using a cut at log ($\lambda_{BHAR}$/erg s$^{-1} M_{star}^{-1}$) $=$ 32.6 
(Figure 5, middle panel);
(b) 251 \textit{low} and 260 \textit{high} SFR AGN cutting at log (SFR/M$_{star}$ yr$^{-1}$)  $=$ 1.4
(Figure 5, lower panel).

In detail, we found a large-scale bias 
b = 2.32$^{+0.11}_{-0.11}$ and b = 2.40$^{+0.10}_{-0.10}$
for the \textit{low} and \textit{high} $\lambda_{BHAR}$, respectively. 
These results suggest no bias evolution with specific BHAR,
with a corresponding typical mass of the hosting halos of log(M$_h$/M$_{\odot}$ h$^{-1}$) = 13.02$^{+0.08}_{-0.08}$ and 13.11$^{+0.06}_{-0.06}$,  respectively. 
On the contrary, AGN with \textit{low} SFR are more 
clustered and reside in more massive dark matter halos 
(log M$_h$/M$_{\odot}$ h$^{-1}$ = 13.14$^{+0.06}_{-0.06}$)
compared to \textit{high} SFR objects (log M$_h$/M$_{\odot}$ h$^{-1}$ = 12.94$^{+0.06}_{-0.06}$).

\begin{figure*}
	\resizebox{\hsize}{!}
	{
		\includegraphics[width=1.2\textwidth]{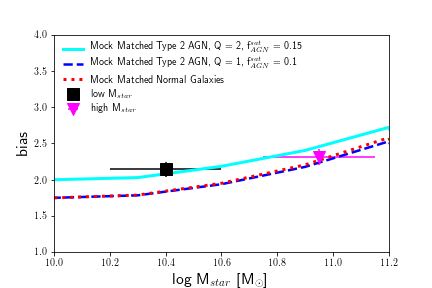}
		\includegraphics[width=1.2\textwidth]{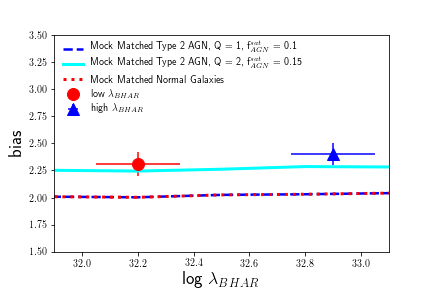}
	}
	\caption{\footnotesize Large-scale bias evolution as a function of host galaxy stellar mass (left panel) and specific BHAR (right panel) for CCL Type 2 AGN and mock AGN matched to have the same host galaxy properties of CCL Type 2 AGN, for $Q$ = 1 and = 2 (according to the legend), which correspond to a relative fraction of satellite AGN $f_{AGN}^{sat}$ = 0.1 and =0.15, respectively. The error bars on x-axis represent the typical error on the stellar mass and specific BHAR estimates in COSMOS. For comparison, the dotted red lines show the large-scale bias as a function of  $M_{star}$ and $\lambda_{BHAR}$ for mock matched normal galaxies.}
	\label{fig:6}
\end{figure*}


\section{\textbf{Discussion}}
\label{sec:conc} 

\subsection{Clustering depence on stellar mass}

We have performed clustering measurements as a function 
of host galaxy stellar mass, specific BHAR and SFR
by using CCL COSMOS Type 2 AGN at mean z $\sim$1.
In particular, our results suggest a constant bias 
evolution as a function of $M_{star}$ for Type 2 AGN 
in the particular stellar mass and redshift range investigated 
in this work (Figure 6). 

As shown in Figure 7, our results are in agreement with the $M_{star}-M_h$ 
relation found in Viitanen et al. (2019) for XMM-COSMOS AGN at similar redshift. 
It is worth noticing that although our study 
includes XMM-COSMOS AGN, we only focused on Type 2 AGN. Moreover, our 
larger CCL COSMOS catalog allows us to split the sample
according to the galaxy stellar mass while having matched SFR, specific BH accretion rate and z distributions. 

Our results are also in agreement with Mountrichas et al. (2019), albeit they 
infer a sligthly steeper $M_{star}-M_h$ relation at mean z$\sim$0.8. 
In detail, they found that high stellar mass XMM-XXL AGN 
(both Type 1 and Type 2 sources) resides in slightly more 
massive halos than low stellar mass objects, when applying a cut at 
log (M$_{star}$/M$_{\odot}$)  = 10.8 (10.7 in our work). They also show that the same $M_{star}$ - $M_h$ relation is observed for a matched sample of normal non active galaxies. 

The relation between galaxies and their hosting halos has been derived for 
normal galaxies in different clustering 
studies (Zheng et al. 2007; Wake et al. 2011, Mostek et al. 2013; Coil et al. 2007) 
at similar  (z $\sim$ 1) and higher redshift (Bielby et al. 2013; Legrand et al. 2018). These results suggest a more steeper $M_{star}$ - $M_h$ relation for non active galaxies than found for CCL Type 2 AGN and from 
previous clustering measurements of AGN and for matched normal galaxies.  

\subsection{Clustering dependence on $\lambda_{BHAR}$ and SFR}

We found a constant typical dark matter halo mass as a function of 
specific BHAR, $\lambda_{BHAR}$. The same trend 
has been observed at similar redshift for XMM-COSMOS AGN by Viitanen et al. (2019) and 
at lower redshift (0.2 $<$ z $<$ 1.2) by Mendez et al. (2016),
using PRIMUS and DEEP2 redshift surveys. 
Assuming a bolometric correction, the specific BHAR can be considered as a proxy 
of the Eddington ratio (Equation 1). Krumpe et al. (2015) found no statistically 
significant clustering dependence on $\lambda_{Edd}$ for RASS Type 1 AGN at 
0.16 $<$ z $<$ 0.36. Our results suggest the same constant bias evolution with
Eddington ratio for both RASS Type 1 and COSMOS Type 2 AGN at different redshifts.


On the contrary, we found a negative bias dependence on SFR (see Figure 7), with lower SFR AGN 
more clustered than higher SFR objects. 
This clustering result suggests that, splitting the sample according to SFR, leads to larger hosting halo mass being associated with lower SFR AGN, i.e. given the same galaxy stellar mass distribution, AGN in low SFR galaxies reside in more massive halos than AGN in high SFR hosts.

AGN clustering analysis as a function of host galaxy SFR has been 
performed by Mountrichas et al. (2019) for X-ray selected XMM-XXL AGN at z = [0.5 - 1.2].
In particular, they split the AGN sample by using a cut at 
log (SFR/M$_{star}$ yr$^{-1}$)  = 1.1 (1.4 in our study) and derived 
typical halo masses of log (M$_h$/M$_{\odot}$ h$^{-1}$) = 13.08$^{+0.38}_{-0.32}$
and 12.54$^{+0.22}_{-0.18}$ at z$\sim$0.8, for the \textit{low} and \textit{high} SFR subsets, respectively.
These results agree very well with our findings for CCL Type 2 AGN at higher redshift,
suggesting a negative clustering dependence as a function of SFR. 

A similar trend has been observed for normal non-active galaxies 
in different studies. Mostek et al. (2013) studied stellar mass-limited samples of DEEP2 galaxies 
and found that the clustering amplitude increases with decreasing SFR. Similarly,
Coil et al. (2007) suggested a strong evolution of the large-scale bias with 
SFR (and sSFR) in PRIMUS and DEEP2 surveys.

Our measured SFR - M$_h$ relation agrees with the environment 
quenching picture, where most galaxies in clusters are passive, 
regardeless of their mass (e.g. Peng et al. 2010, 2012).
In particular, the passive fraction in both central and satellite galaxies 
strongly correlates with the halo mass at fixed stellar mass. 
Above a characteristic halo mass ($\sim10^{12}$ M$_{\odot}$) 
cooling times are long, and the gas that accretes onto the
galaxy is hot, so star formation is inefficient (e.g. Gabor \& Dave 2015, 
Birnboim \& Dekel 2003, see also Peng, Maiolino \& Cochrane 2015, for observational
evidence for quenching via gas-exhaustion, or ‘strangulation’).
However, CCL Type 2 AGN also cover the redshift range 
in which a difference in galaxy SFR properties with environments is ceasing (e.g. George et al. 2013, Erfanianfar et al 2016 at $z \la 1.2$) and the main difference being the fraction of bulge-dominated galaxies as a function of halo mass.  Then the SFR cut might also lead to higher weight of bulge-dominated galaxies, which reside in more massive halos.

%
%
%
%
%

\subsection{Comparison with AGN mock catalogs}

\subsubsection{Methodology}

Several semi-analytical models and hydrodynamical simulations 
(e.g. Springel et al. 2005; Hopkins et al. 2006; Menci et al. 2008) 
have been developed in recent years to describe the main mechanisms 
that fuel the central BHs. With suitable adjustment of parameters,
these models can explain many aspects of AGN phenomenology 
(e.g. Hopkins et al. 2006, 2008).
However, our scant knowledge of the key processes imposes a 
heavy parameterization of the physics
regulating the cooling, star formation, feedback, and merging of baryons. 
Thus, current state-of-the
art models present serious degeneracies, i.e. they reproduce 
similar observables by invoking very
different scenarios (Lapi et al., 2011). Instead, in Semi-Empirical Models (SEMs) 
variables (such as galaxy stellar mass, BH mass, AGN luminosity) 
are assigned through a combination of observational
and theoretical scaling relations. SEMs represent an original 
and competitive methodology, which is fast, flexible and 
relies on just a few input assumptions. Mock catalogs 
of galaxies and their BHs can be created via semi-empirical 
relations starting from large samples of dark matter halos extracted
from N-body simulations. Therefore, AGN mock catalogs effectively 
provide a complementary approach to more complex models 
of AGN and galaxy evolution (e.g. Conroy \& White 2013) which can be directly compared
with current and future clustering measurements.
The semi-empirical model for the large-scale distribution of AGN 
has been used to make realistic predictions for the clustering signal in 
future experiments and test observational selection effects and biases 
(e.g. Georgakakis et al. 2018, Comparat et al. 2019).

In this study, we compared our observational results in COSMOS with mock catalogs
of active and non-active galaxies created via SEMs 
based on large N-body simulations. The full description of numerical routines 
to create mock catalogs of galaxies and their BHs by using SEMs 
is given in Allevato et al. (in preparation). Here we only describe the important 
steps in the generation of the AGN mock catalogs and we refer the reader to Allevato et al. 
for more details.
\setcounter{footnote}{0}
First, we extracted a large catalog of dark matter halos and subhalos from MultiDark\footnote{www.cosmosim.org}-Planck 2
(MDPL2, Riebe et al. 2013) at the redshift of interest, which currently provides the largest publicly available set of high-resolution and large volume N-body simulations (box size of 1000 h$^{-1}$Mpc, mass resolution of 1.51$\times$10$^9$ h$^{-1}$M$_{\odot}$). 
The \texttt{ROCKSTAR} halo finder (Behroozi et al. 2013) has been 
applied to the MDPL2 simulations to identify halos and flag those 
(sub-halos) that lie within the virial radius of a more massive host halo. 
The mass of the dark matter halo is defined as the virial mass in the case of 
host halos and the infall progenitor virial mass for sub-halos. 
In the analysis that follows, we use the simulation snapshot at 
z = 1, which corresponds to the mean redshift of our AGN 
clustering measurements.

We assign to each halo: 
(a) a galaxy stellar mass
deduced from the Grylls et al. (2019) semi-empirical relation, inclusive of intrinsic (0.15 dex) and measurement scatter (0.2 dex for stellar mass estimates in our COSMOS sample). Satellites are assigned a stellar mass at the redshift of infall; 
(b) a BH mass assuming empirical BH-galaxy mass relation derived in Shankar et al. (2016), inclusive of a stellar mass dependent scatter;
(c) an X-ray luminosity following the observationally 
deduced specific BHAR distribution described by a Schechter function as suggest in Bongiorno et al. (2012, 2016), Aird et al. (2016), Georgakakis et al. (2017);
(d) a SFR following the SFR - stellar mass relation described in Tomczak et al. (2016) for main sequence galaxies, including intrinsic (0.2 dex) and measurement (0.2 dex for our  COSMOS sample) scatter;
(e) an hydrogen column density $N_H$ assigned following the Ueda et al. (2014)
empirical distribution such that AGN can be classified into Type 2 obscured, Type 1 unobscured 
and Compton Thick AGN;
(f) a duty cycle, i.e. a probability for each BH of being active, following Schulze et al. (2015).

The main difference between our approach and recent studies based on semi-empirical models
(e.g. Georgakakis et al. 2019, Comparat et al. 2019) is that the latter assign an Eddington ratio to each mock BH, and consider \textit{active} the objects with a specific BHAR above a given value.
On the contrary, we assign to each BH an Eddington ratio combined with a probability of being active, which depends on the BH mass. A comparison between the different semi-empirical models is beyond the scope of this paper and is discussed in Allevato et al. in prep.

We then selected a sample of mock AGN, matched to have the same galaxy stellar mass, specific BHAR, SFR and X-ray luminosity distributions of CCL Type 2 AGN, including only Type 2 objects selected according to $N_H$ values. This corresponds to mock AGN living in dark matter halos (parent halos for satellite galaxies) with logM$_h[M_{\odot}]>12$. To each parent halo mass $M_{halo}$, a bias is assigned which satisfies b = b($M_{halo}$), following the same bias-mass relation used for CCL Type 2 AGN (van den Bosch 2002 and Sheth et al. 2001). 
Similarly, we selected a sample of normal non-active galaxies with the same galaxy stellar mass, SFR and X-ray luminosity distributions of CCL Type 2 AGN.

\begin{figure*}
	\resizebox{\hsize}{!}
	{
		\includegraphics[height=3.6\textwidth]{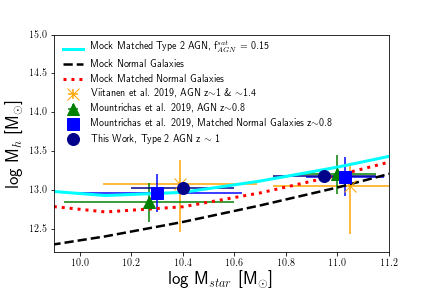}
		\includegraphics[height=3.6\textwidth]{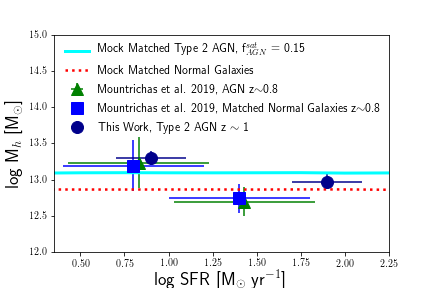}
	}
	\caption{\footnotesize Galaxy stellar mass as a function of dark matter halo mass as derived for CCL Type 2 AGN (dark blue circles) at mean z$\sim$1, XMM-COSMOS AGN (orange stars) at z$\sim$1 (low $M_{star}$) and $\sim$1.4 (high $M_{star}$) in Viitanen et al. (2019), XMM-XXL AGN (green triangles) and matched normal galaxies (blue squares) in Mountrichas et al. (2019) at mean z$\sim$0.8. The error bars on x-axis represent typical measurement error on the stellar mass estimates of each subsample. The $M_{star}-M_h$ and $M_{star}$ - SFR relations for matched mock AGN (continuous grey line) and mock matched normal galaxies (dotted red line) are shown. For comparison, the halo-stellar mass relation is shown for the full sample of mock galaxies as dashed black line.}
	\label{fig:7}
\end{figure*}

\subsubsection{\textbf{Bias for mock AGN}}

We follow the formalism of Shankar et al. (submitted) to derive the bias of mock AGN and normal galaxies as a function of galaxy stellar mass, specific BHAR and SFR, by using the relative probabilities of central and satellite galaxies of being active $Q = U_s/U_c$. \\
The bias of mock objects with stellar mass in the range $M_{star}$ and $M_{star} + dM_{star}$ is estimated as:
\begin{equation}
b(M_{star}) = \frac{1}{N_{bin}} \sum_{i=1}^{N_{bin}} b[M_{h,i} (M_{star})] 
\end{equation}
where for each bin of stellar mass the sum runs over all host parent halos. 
If the  probabilities for galaxies to be active, i.e. the AGN duty cycle $U(M_{star}) = U_{c}(M_{star})+U_{s}(M_{star})$ is included, the generalised formula is then:
\begin{equation}
b = \frac{  \left[ \sum_{i=1}^{N_{c}} U_{c,i}(M_{star}) b_{c,i}(M_{star}) + \sum_{i=1}^{N_{s}} U_{s,i}(M_{star}) b_{sat,i}(M_{star}) \right] }           {   \left[ \sum_{i=1}^{N_{cen}} U_{c,i}(M_{star})  + \sum_{i=1}^{N_{s}} U_{s,i}(M_{star}) \right]  } 
\end{equation}
where $U_{c}(M_{star}) = U(M_{star})N(M_{star})/(N_c(M_{star}) + Q + N_s(M_{star})$ is the duty cycle of central AGN, $U_{s}(M_{star}) = QU_{c}(M_{star})$ is the duty cycle of satellite AGN
and $N(M_{star}) = N_c(M_{star}) + N_s(M_{star})$ is the number of central and satellite galaxies, 
in the stellar mass bin $M_{star}$ and 
$M_{star} + dM_{star}$.

When $U_s = U_c$, i.e. if all central and satellite galaxies are active or share equal probabilities of being active, Equation 10 reduces to Equation 9. 
It is important to notice that the bias is thus mainly affected by $Q$ and
then by the fraction of AGN in satellite halos $f_{sat}^{AGN}$, so that
$Q   = f_{sat}^{AGN} (1-f_{sat}^{BH}) / {1-f_{sat}^{AGN}}f_{sat}^{BH} $,
where $f_{sat}^{BH}=N_s/(N_s+N_c)$ is the total fraction of (active and non active) BHs in satellites with host galaxy stellar mass within  $M_{star}$ and $M_{star} + dM_{star}$.
In our mock of AGN matched to CCL Type 2 AGN, we have a total fraction of satellite galaxies $f_{sat}^{BH} \sim$ 0.1, which corresponds to $Q=1$ for a fraction of satellite AGN $f_{sat}^{AGN}$ $\sim$ 0.1 and $Q=2$ for $f_{sat}^{AGN}$ $\sim$ 0.15, respectively.

Similarly, Equation 10 can be written in bins of specific BHAR and SFR, for $Q$ = 1 and 2, respectively.


\subsubsection{Predictions from SEMs}

Figure 6 shows the mean bias as a function of the host galaxy stellar mass and specific BHAR 
for mock AGN (Equation 10) and for normal galaxies (Equation 9), when $Q$ = 1 and  = 2, which corresponds 
to a fraction of satellite AGN $f_{AGN}^{sat} \sim$ 0.1 and $\sim$ 0.15, respectively.  
The host galaxy stellar masses in the mock catalog are assigned by following Grylls et al. (2019), i.e. are defined as Sersic + exponential model (Bernardi et al. 2013)  with a mass-to-light ratio from Bell et al. (2013). To correct for the different definition in COSMOS (Bruzual \& Charlot 2013, Chabrier 2003) we need to decrease the stellar masses of mock objects by a factor of $\sim$0.15 dex (Grylls et al., submitted). 

The bias - $M_{star}$ relation for mock AGN and matched mock normal galaxies is slightly steeper compared to clustering measurements of COSMOS AGN, when we assume the same probability of satellite and central galaxies of being active ($Q$=1).
In particular, we found that CCL Type 2 AGN host galaxies with low $M_{star}$ reside in 
slightly more massive halos than mock AGN and normal galaxies of similar stellar mass. 
It is important to notice that when $Q$ = 1, Equation 10 reduces to the simple case of Equation 9, and AGN and normal galaxies have the same bias - $M_{star}$ relation.

Our results for CCL Type 2 AGN can be reproduced if we assume 
a larger satellite AGN fraction $f_{AGN}^{sat} \sim 0.15$ and $Q \sim$ 2.
For instance, Leauthaud et al. (2015) found $f_{AGN}^{sat} \sim$ 0.18 for COSMOS AGN at z $\la$1. 
The importance of the relative fraction of satellite AGN has been underlined also in Viitanen et al. (2019). In fact, they suggest that when excluding AGN that are  
associated with galaxy groups, the bias of low $M_{star}$ objects decreases,
while not affecting the high stellar mass systems. Similar results are found
in Mountrichas et al. (2013) for moderate luminosity X-ray AGN at z$\sim$1.
This is due to the fact that at lower stellar mass, AGN are more likely in satellite galaxies hosted 
by more biased and massive parent dark matter halos. Similarly, the bias as a function of specific BHAR and SFR of CCL Type 2 AGN is better reproduced by mock AGN with $Q$ = 2.
It is worth noticing that the bias - $M_{star}$, bias - $\lambda_{Edd}$ and bias -SFR relations of normal non active mock galaxies are independent of the relative duty cycle of AGN in satellite and central halos, $Q$. 

Our results thus suggest that for COSMOS Type 2 AGN at z$\sim$1, the relative probabilities of AGN in satellites is $\sim$ 2 times larger than in central halos, with a fraction of AGN in satellite halos consistent with  $f_{AGN}^{sat} \sim 0.15$. 
Starikova et al. (2011) studied the HOD of AGN detected by the Chandra X-Ray Observatory in the Bootes field over a redshift interval z = [0.17-3],
showing a satellite fraction of $\sim$10\%.
Allevato et al. (2012) performed direct measurement of the HOD for COSMOS AGN 
based on the mass function of galaxy groups hosting AGN and found that the duty cycle of satellite AGN is comparable
or even larger than that of central AGN, i.e. $Q  \ga$ 1.
The central locations of the quasar host galaxies are expected in
major merger models because mergers of equally sized galaxies
preferentially occur at the centers of DM halos (Hopkins et al.
2008). Our predictions from mock matched AGN suggests a high fraction of satellite AGN, in agreement with studies that find a small fraction of AGN associated
with morphologically disturbed galaxies (Cisternas et al. 2011, Schawinski
et al. 2011, Rosario et al. 2011) and that suggest secular processes and bar instabilities being efficient in producing luminous AGN (e.g. Georgakakis et al. 2009, Allevato et al. 2011).



Figure 7 shows the typical dark matter halo mass as a function of 
galaxy stellar mass, as found for CCL Type 2 AGN at mean 
z$\sim$1 (black circles), for XMM-XXL AGN at z$\sim$0.8 (Mountrichas et al. 2019, green triangles) 
and as predicted for matched mock AGN (continuous line) and matched normal 
galaxies (dotted line) for $Q$ = 2.
The predictions from mock AGN 
well reproduce the observations in COSMOS,
as well as the results for XMM-XXL AGN 
at similar redshift.

A sligthly steeper trend is observed for matched normal mock galaxies, with non active BH residing in less massive parent halos than mock AGN with low host galaxy stellar mass.
This is in contrast with the results of 
Mountrichas et al. (2019), at least for low $M_{star}$ 
galaxies. In fact, they suggest that AGN have the same stellar-to-halo mass ratio of matched 
normal galaxies, at all stellar masses. 
Previous clustering studies of normal galaxies (e.g. Zheng et al. 2007, Coil et al. 2007) suggest a steeper halo - stellar mass relation at similar 
redshift. A flatter $M_{star}$ - M$_h$ relation found for mock matched galaxies is mainly due: (i) to the large scatter (measurement 0.2 dex and intrinsic 0.15 dex) in the input relation (see Sec 4.3.1) used to create the mock catalog, in order to reproduce the stellar mass measurement error in COSMOS; and (ii) to the selections in terms of L$_X$, $M_{star}$, specific BHAR and SFR applied to match CCL Type AGN hosts.
In fact, Figure 7 shows that
when selecting a subsample of 
galaxies matched to have the same properties of CCL Type 2 AGN hosts (dotted red line), the galaxy bias is driven up at low stellar mass with respect to the full galaxy population (dashed line).

Figure 8 shows the halo mass - SFR relation for 
mock AGN and matched normal galaxies. 
The predictions are almost
consistent with our results in COSMOS and with previous studies in XMM-XXL at similar redshifts 
(Mountrichas et al. 2019), but suggest a slightly flatter SFR - M$_h$ relation than observed.
A constant SFR as a function of the halo mass is a consequence of the almost flat M$_{star}$ - M$_{h}$ relation obtained for mock objects, combined with the input assumption that 
each mock AGN and galaxy follow a simple main sequence SFR-M$_{star}$ relation.

It is important to notice that 
given the limited sample of CCL Type 2 AGN in bins of M$_{star}$, SFR and specific BHAR, we can only estimate typical halo masses as a function of AGN host galaxy properties, from the modelling of the 2-halo term. The full halo mass 
distribution and halo occupation (possibly separating the contribution of AGN in central and satellite galaxies) require higher statistics to constrain the clustering signal at small scale. Currently, one possibility to overcome the low statistics is to combine available samples of AGN in X-ray surveys, like \textit{Chandra}-COSMOS Legacy (Civano et al. 2016), AEGIS and 4Ms CDFS (Georgakakis et al. 2015) and XMM-XXL (Mendez et al. 2016), with robust host galaxy property estimates. 
Following this approach the number of AGN with known spectroscopic 
redshift can be almost doubled with respect to the sample of Type 2 AGN used in this work. During next years, the synergy of eRosita, 4MOST, WISE and Euclid, JWST in the near future, will 
allow us to derive host galaxy stellar mass and SFR estimates of millions of moderate-high luminosity AGN up to z$\sim$2.

Moreover, our clustering measurements refer to Type 2 AGN only. New 
\textit{Chandra}-COSMOS Legacy are now available for Type 1 AGN (Suh et al. 2019) and then AGN clustering dependence on host galaxy properties will be probed in the near future as a function of obscuration. 

\section{Conclusions}

In this paper, we have performed clustering measurents of CCL COSMOS  Type 2 AGN at mean z$\sim$1, to probe the AGN large-scale bias 
dependence on host galaxy properties, such as galaxy stellar mass,
specific BHAR and SFR.
Our main findings can be summarized as follow:
\begin{itemize}
	\item We found no dependence of the AGN large-scale
	bias on galaxy stellar mass and specific BHAR, suggesting almost flat   
	$M_{star}$ - M$_h$ and $\lambda_{BHAR}$ - M$_h$ relations.
	\item We found a negative 
	clustering dependence on SFR, with the typical hosting
	halo mass increasing with decreasing SFR;
	\item Mock catalogs of AGN matched to have the same 
	host galaxy properties of COSMOS Type 2 AGN predict the observed 
	$M_{star}-M_h$, SFR-M$_h$ and $\lambda_{BHAR}-M_h$ relations, when assuming a fraction of satellite AGN $f_{AGN}^{sat} \sim$ 15\% and then $Q$ = 2; 
	\item Mock matched normal galaxies follow a slightly steeper $M_{star}-M_h$ relation, with \textit{low} mass mock galaxies residing in slightly less massive halos than mock AGN of similar mass. Similarly, mock galaxies reside in less massive hosting halos than mock AGN with similar specific BHAR and SFR, at least for $Q>$ 1.
\end{itemize}

\begin{acknowledgements}

VA acknowledges funding from the European Union's Horizon 2020 research and innovation programme under grant agreement No 749348. TM is supported by CONACyT 252531 and UNAM-DGAPA PAPIIT IN111319.

\end{acknowledgements}

\clearpage
%
%
%

\end{document}